\documentclass{cimento}

\usepackage{graphicx}  

\title{A set of generalized parton distributions }   
\author{P.~Kroll \from{ins:x}  \ETC,}
\instlist{\inst{ins:x} 
Fachbereich Physik, Universit\"at Wuppertal, D-42097 Wuppertal,
Germany and Institut f\"ur Theoretische Physik, Universit\"at
    Regensburg, D-93040 Regensburg, Germany}

\PACSes{\PACSit{13.60Le}{13.60Fz}{12.39St}}

\begin{document}

\maketitle

\begin{abstract}
The information about generalized parton distributions (GPDs) extracted from exclusive meson 
leptoproduction (DVMP) within the handbag approach is summarized. Details are only discussed
for the GPD $E$ and the transversity ones. It is also commented on results for deep virtual 
Compton scattering (DVCS) evaluated from these GPDs. 
\end{abstract}

\section{Introduction}
The handbag approach to hard exclusive leptoproduction of photons and 
mesons off protons has extensively been studied during the last fifteen
years. This approach is based on factorization of the process amplitudes 
in hard subprocesses, e.g.\ $\gamma^*q\to \gamma (M) q$, and soft hadronic matrix 
elements parametrized in terms of GPDs. This factorization property has been 
shown to hold rigorously  in the generalized Bjorken regime of large photon 
virtuality, $Q$, and large energy $W$ but fixed $x_B$. Since most of the data, in 
particular those from the present Jlab, are not measured in this kinematical regime one 
has to be aware of power corrections from various sources. Which kind of power 
correction is the most important one and is to be taken into account is still
under debate. Nevertheless progress has been made in the understanding of the DVCS 
and DVMP data. In this talk, presented at QCD'N12 and SPIN12 (see \cite{kroll}), I 
am going to report on an extraction of the GPDs from 
DVMP \cite{GK1}. In this analysis the GPDs are constructed from double 
distributions (DDs)\cite{mueller94, rad98} where the latter are parametrized as
zero-skewness GPDs times weight functions which generate their skewness dependence.
The ans\"atze for the zero-skewness GPDs consist of their corresponding forward limits 
multiplied by exponentials in Mandelstam $t$ with profile functions parametrized
in a Regge-like manner with slopes of appropriate Regge trajectories and constants
for the $t$ dependence of their residues. These profile functions are simplified versions 
of more complicated ones proposed in \cite{DFJK4}. At small momentum 
fractions, $x$, they fall together with the ones used in \cite{DFJK4}. Because of a strong 
$x - t$ correlation observed in \cite{DFJK4} the Regge-like profile functions can 
only be applied at small $-t$. The forward limits of the zero-skewness GPDs are in 
some cases ($H$, $\widetilde{H}$, $H_T$) given by the usual parton densities, in 
other cases (for the $E$-type ones) they are parametrized like the parton densities 
with a number of free parameters adjusted to experiment. 

In Sect.\ 2 some details on the extraction of the GPDs are presented. In Sects.\ 3 $E$ 
is discussed and in Sect.\ 4 the transversity GPDs. A summary is given in Sect.\ 5. 

\section{Extraction of the GPDs from hard meson leptoproduction}
As an example we quote the convolution formula for the production of longitudinally 
polarized vector mesons:
\begin{equation}
{\cal F}_{\rm V}(\xi, t, Q^2) = \sum_{i,\lambda} \int^1_{x_i} dx 
            {\cal A}^i_{0\lambda, 0\lambda}(x,\xi,Q^2,t=0)\, F^i(x,\xi,t)
\label{eq:convolutions-mesons}
\end{equation}
where $i=g, q$, $x_g=0$, $x_q=-1$ and $F$ either $H$ or $E$. Similar
convolution formulas hold for transversely polarized vector mesons and for pseudoscalar 
mesons. The subprocess amplitude ${\cal A}$ for partonic helicity $\lambda$ is to be 
calculated perturbatively using $k_\perp$-factorization. This means that in the subprocess
quark transverse degrees of freedom as well as Sudakov suppression \cite{li} are taken 
into account. The emission and reabsorption of the partons by the protons are treated 
collinearly. This approach also allows to calculate the amplitudes for transversely 
polarized photons and like-wise polarized vector mesons which are infrared singular in 
collinear factorization. The transverse photon amplitudes are rather strong for 
$Q^2\leq 10\,{\rm GeV}^2$ as is known from the ratio of longitudinal and transverse cross 
sections for $\rho^0$ and $\phi$ production \cite{h1}. The approach used in \cite{GK1} 
bears similarity to the color dipole model \cite{frankfurt}.

There is another problem with vector meson production: In collinear factorization the cross 
section for the production of $\rho^0$ drops as $1/Q^6(\log{Q^2})^n$  with increasing $Q^2$ 
while experimentally \cite{h1} it approximately falls as $1/Q^4$. In the above sketched
approach the required suppression of the amplitudes at low $Q^2$ is generated by the evolution
of the GPDs and by $k_\perp/Q$ effects. In \cite{MM} however GPDs are proposed which have a 
much stronger evolution than those used in \cite{GK1}. At least for HERA kinematics the GPDs 
advocated for in \cite{MM} lead to fair fits of the HERA data on DVMP in collinear factorization. 

In \cite{GK1} parameters of the DDs are fitted to the available data on 
$\rho^0$, $\phi$ and $\pi^+$ production from HERMES, COMPASS, E665, H1 and ZEUS. 
The data cover a large kinematical range: $3\,{\rm GeV}^2\leq Q^2\leq 100\,{\rm GeV}^2$, 
$4\,{\rm GeV}\leq W\leq 180\,{\rm GeV}$, i.e. Bjorken-$x$ and, hence, skewness, is small.
Data from the present Jlab (characterized by large $x_B$ and small $W$) are not taken into 
account in these fits because they are likely affected by strong power corrections 
at least in some cases (e.g. $\rho^0$ production). Constraints from nucleon form
factors and from positivity bounds \cite{DFJK4} are taken into account.
The analysis is strongly simplified by the fact that, for small $x_B$, the $\rho^0$ and 
$\phi$ cross sections are under control of contributions from the GPD $H$, other GPDs can
be ignored. Since $H$ is rather well fixed by many constraints (PDFs, nucleon form factors) 
the vector meson cross sections allow to pin down the remaining few free parameters of $H$. 
All other GPDs are much less well known than $H$. Their extraction requires polarization
observables or hard leptoproduction of pseudoscalar mesons. The latter process is however 
complicated to analyze since many GPDs contribute at the same level. The analysis performed
in \cite{GK1} leads to a fair description of all the mentioned data. What has been learned 
about the GPDs from this analysis is summarized in Tab.\ \ref{tab1}. For details of the 
parametrization and values of the parameters it is referred to \cite{GK1,gk6,kms}. 

In \cite{kms} the GPDs extracted in \cite{GK1} have been exploited to compute DVCS to 
leading-twist accuracy and leading-order of pQCD  while the Bethe-Heitler contribution is 
worked out without any approximation. It should be realized that, to this level of accuracy,
collinear emission and reabsorption of the partons from the protons forces the partonic 
subprocess of DVCS to be collinear as well. A detailed comparison of this theoretical 
approach with experiment performed in \cite{kms}, reveals reasonable agreement with 
HERMES, H1 and ZEUS data and a less satisfactory description of the large-skewness, 
small $W$ Jlab data (see talk by F. Sabatie this conference). Note that the GPDs extracted 
in \cite{GK1} are not optimized for the latter kinematical region. It should also be mentioned 
that in the same spirit a DVCS analysis is performed in \cite{MM,kmls}. 

\begin{table}[t]
\renewcommand{\arraystretch}{1.2} 
\begin{center}
\begin{tabular}{| c || c | c | c |}
\hline   
GPD &  probed by &  constraints &  status \\[0.2em]
\hline
$H$(val) & {\small $\rho^0, \phi$ cross sections} & {\small PDFs, Dirac ff} & {\small ***} \\[0.2em]
$H$(g,sea) & {\small $\rho^0, \phi$ cross sections} & {\small PDFs} & {\small ***} \\[0.2em]
$E$(val)& {\small $A_{UT}(\rho^0, \phi)$} & {\small Pauli ff} & {\small **} \\[0.2em]
$E$(g,sea) &  - & {\small sum rule for $2^{nd}$ moments}
& {\small {-}}\\[0.2em]
$\widetilde{H}$ (val)  & {\small $\pi^+$ data} &{\small pol.\ PDFs, axial ff} &
{\small **} \\[0.2em]
$\widetilde{H}$(g,sea)  & $A_{LL}(\rho^0)$  & {\small polarized PDFs} & {\small
    *}\\[0.2em]
$\widetilde{E}$ (val)& {\small $\pi^+$ data} & pseudoscalar ff  
                                          & {\small *} \\[0.2em]
$H_T, \bar{E}_T$(val) & {\small $\pi^+$ data} & {\small transversity PDFs} 
                                          & {\small *}\\[0.2em]
\hline
\end{tabular}
\caption{\label{tab1} Status of small-skewness GPDs as extracted DVMP. No information 
is presently available on GPDs not appearing in the table. Except of $H$ for gluons 
and sea quarks all GPDs are only probed for scales of about $4\,{\rm GeV}^2$. For comparison 
five stars are assigned to PDFs.} 
\end{center}
\renewcommand{\arraystretch}{1.0}   
\end{table}

\section{The GPD $E$}
Let me now discuss the GPD $E$ in some detail. The analysis of the nucleon form 
factors carried through in \cite{DFJK4} provided the zero-skewness GPDs for valence 
quarks which can be used to construct the DDs. Since in 2004 data 
on the neutron form factors were only available for $-t\leq 2\,{\rm GeV}^2$ the 
parameters of the zero-skewness GPD $E^q_v$ were not well fixed; a wide range of 
values were allowed for the powers $\beta^u_e$ and $\beta^d_e$  which control the 
large-$x$ behavior of the forward limits of $E^q_v$. In the recent 
reanalysis of the form factors \cite{dk12} making use of all new data which for 
the neutron now extend to much larger values of $t$, similar results for the 
valence-quark GPDs are obtained but the powers $\beta^q_e$ are now better determined.

Not much is known about $E^g$ and $E^{\rm sea}$. There is only a sum rule for the 
second moments of $E$ \cite{diehl-kugler} at $t=\xi=0$
\begin{equation}
\int_0^1 dx E^g(x,\xi=0,t=0) = e_{20}^g =-\sum e_{20}^{q_v} -2\sum e_{20}^{\bar{q}}\,.
\label{eq:SumRuleE}
\end{equation} 
It turns out that the valence contribution to the sum rule is very small 
\cite{DFJK4,dk12}. Hence, the second moments of the gluon and sea-quark GPD $E$ 
cancel each other almost completely. For parametrizations of the forward limits of 
$E$ which do not have nodes except at the end-points (see e.g.\ \cite{GK4}) this 
property approximately holds of other moments as well and even for convolutions like 
(\ref{eq:convolutions-mesons}). For $E^s$ there is also a positivity  bound for its 
Fourier transform with respect to the momentum transfer \cite{DFJK4,diehl-kugler,GK4} 
which forbids a large strange quark contribution and, assuming a flavor-symmetric sea, 
a large gluon contribution too. Determining the normalization of $E^s$ by assuming that 
the bound for it is saturated for some values of $x$ (note the bound is quadratic in 
$e_s$), one can subsequently fix the normalization of $E^g$ from the sum rule 
(\ref{eq:SumRuleE}) \cite{GK4}. 

\begin{figure}[t]
\begin{center}
\includegraphics[width=0.37\textwidth, bb=254 25 503 282, clip=true]{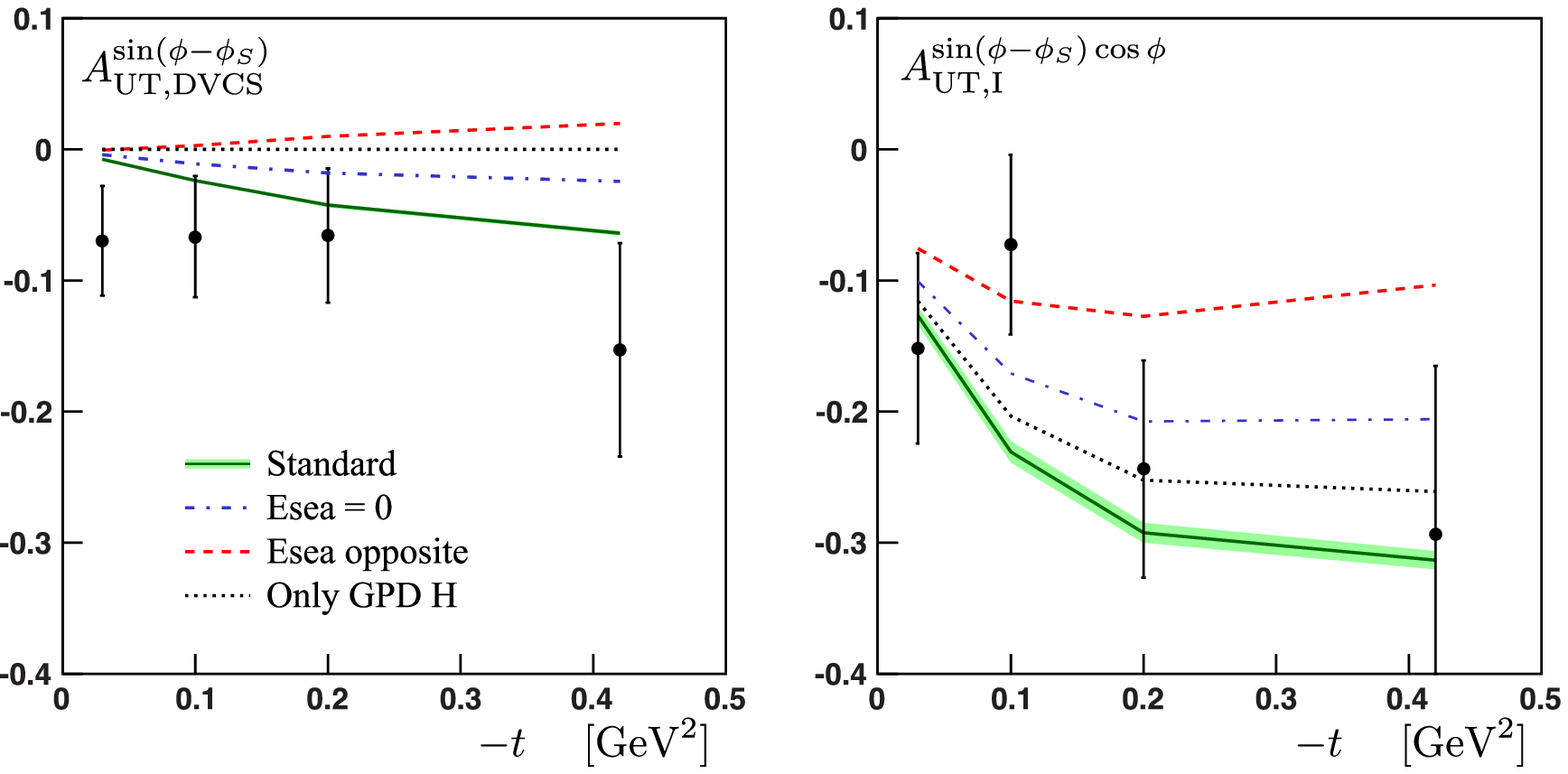}
\hspace*{0.03\textwidth}
\includegraphics[width=0.45\textwidth]{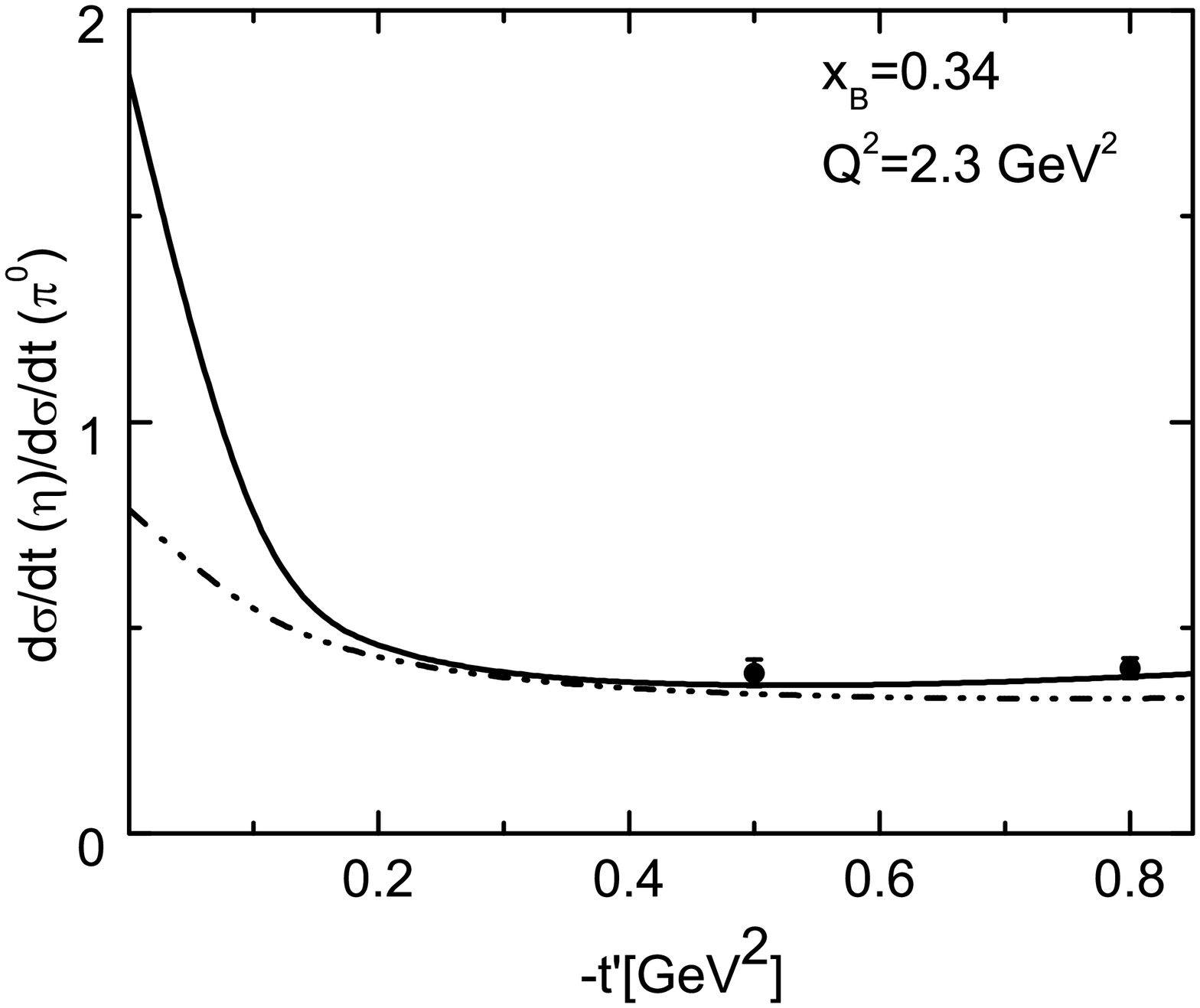}
\caption{Left: The BH-DVCS interference. Data are taken from \cite{hermes-aut-dvcs}, theoretical 
results from \cite{kms}. Right: The ratio of the $\eta$ and $\pi^0$ cross sections versus
$t'$. Preliminary data are taken from \cite{valery}.} 
\label{fig:AUT}
\end{center}
\end{figure}

For given $H$, as for instance extracted from the DVMP cross sections \cite{GK1}, the 
GPD $E$ is probed by the transverse target asymmetry 
\begin{equation} 
 A_{UT} \sim {\rm Im}\Big[{\mathcal E}^* {\mathcal H}\Big]\,.
\end{equation}
The data on $\rho^0$ production from HERMES \cite{hermes-aut} and COMPASS \cite{compass-aut} 
are well fitted by the described parametrization of $E$. However, only $E$ for valence 
quarks matters for $A_{UT}(\rho^0)$ since the sea and gluon contribution to $E$ cancel to 
a large extent as remarked above. Fortunately the analysis of DVCS data \cite{kms} provides 
additional although not very precise information on $E^{\rm sea}$. To leading-order of pQCD 
there is no gluon contribution in DVCS and therefore $E^{\rm sea}$ becomes visible. The 
HERMES collaboration has measured the transverse target asymmetries for DVCS and for the 
BH-DVCS interference term \cite{hermes-aut-dvcs}. Despite the large experimental errors it 
seems that a negative $E^{\rm sea}$ is favored. As an example the data on the BH-DVCS 
interference are shown in Fig.\ \ref{fig:AUT} and compared to the results obtained in \cite{kms}. 
Independent information on $E^g$ would be of interest. This may be obtained from a 
measurement of the transverse target polarization in $J/\Psi$ photoproduction \cite{koempel}. 

The knowledge of $E$ which is admittedly poor, allows for an estimate of the angular 
momenta the partons inside the proton carry. At $\xi=t=0$ they are given by the second moments
of $H$ and $E$
\begin{equation}
2J^a=\Big[q_{20}+e_{20}^q\Big]\,, \qquad 2J^g=\Big[g_{20}+e_{20}^g\Big]\,.
\end{equation}
The values of the $H$-moments can be evaluated from the PDFs, for instance from \cite{cteq6}.
Since a negative $E^s$ is favored as we learned from the combined analysis of DVMP and DVCS, 
$E^g$ is positive (remember the no node assumption). Therefore, the second moment of the 
latter, $e^g_{20}$, is positive and adds to $g_{20}$ which is large and positive as is known 
for a long time (it represents the fraction of the proton momentum carried by gluons). From 
these considerations it follows that the total angular momentum carried by the gluons is 
large as well. According to \cite{GK4}, it amounts to $J^g=0.21-0.29$.   

\section{The transversity GPDs}
There is a second set of four GPDs, the transversity ones which are characterized
by opposite helicities of the emitted and reabsorbed partons. In general they play
a minor role in exclusive reactions and not many phenomenological studies are devoted to 
them (an example is \cite{pire}). However, it became evident recently that the 
transversity GPDs contribute strongly to leptoproduction of pseudoscalar mesons 
\cite{gk6,gk5,liuti}. The first experimental evidence for transversity came from 
the $\sin \phi_s$ harmonics of the $\pi^+$ production cross section measured with a
transversely polarized target \cite{hermes-aut-pi}. From these data we learned that
this observable is large and does not seem to vanish for forward scattering. Such a behavior 
requires a strong helicity non-flip amplitude for transversely polarized virtual photons.
Within the handbag approach this amplitude is under control of the transversity GPD $H_T$
in combination with a twist-3 pion wave function \cite{gk5}. This amplitude is parametrically
suppressed by $\mu_\pi/Q$ as compared to the asymptotically dominant amplitude for
longitudinal polarized photons. Here, $\mu_\pi=m_\pi^2/(m_u+m_d)\simeq 2\,{\rm GeV}$ at the scale
of $2\,{\rm GeV}$ where $m_q$ is a current quark mass. Hence, this twist-3 effect is quite
large for experimentally accessible values of $Q$. Moreover lattice QCD \cite{haegler} provides
some evidence of a large GPD $\bar{E}_T=2\widetilde{H}_T+E_T$ with the same sign and
almost the same size for $\bar{E}^u_T$ and $\bar{E}^d_T$. Both the GPDs, $H_T$ and $\bar{E}_T$ are 
parametrized in an analog fashion than the other GPDs and their parameters are fixed
by fits to the HERMES $\pi^+$ data \cite{hermes-aut-pi,hermes-pip} and by taking recourse to 
the lattice-QCD results.
With the GPDs determined this way, the behavior of the transverse target asymmetry
can be understood quantitatively and predictions for the productions of the $\pi^0$
and other pseudoscalar mesons have been given \cite{gk6}. It turns out that the $\pi^0$
cross section is dominated by the contributions from the transversity GPDs except in the near
forward region; the ratio of the longitudinal and transverse cross sections is much smaller
than 1 for $Q^2\leq 10\,{\rm GeV}^2$. The results for $\pi^0$ production are in fair
agreement with the large Bjorken-$x$ (small $W$) data from CLAS \cite{clas}.   
Another interesting prediction is that the ratio of the $\eta$ and $\pi^0$ cross sections
is much smaller than 1 (except in the near forward region) in sharp contrast to expectations
\cite{strikman}. Also this result which is shown in Fig.\ \ref{fig:AUT}, is in reasonable
agreement with preliminary CLAS data \cite{valery}.
       
\section{Summary}
I have briefly summarized the recent progress in the analysis for hard exclusive
leptoproduction of mesons and photons within the handbag approach. We learned
that the data on both reactions are consistent with each other in so far as they
can be described with a common set of GPDs. In fact the GPDs constructed from
double distributions and with parameters adjusted to the meson data allow for
a parameter-free calculation of DVCS.

Of course the GPDs are not perfect, they are an approximation. Improvements are required
for which the future COMPASS and Jlab12 will be of help. Possible improvements
may include the use of more recent versions of the PDFs, eventual modifications
of the parametrizations of the GPDs, in particular of the profile functions of the
corresponding double distributions, and allowance for a non-zero $D$-term. Updated
zero-skewness GPDs $H$ and $E$ for valence quarks have already been obtained from the recent
analysis of the nucleon form factors \cite{dk12}. These result are not yet used in 
evaluations of the DVCS and meson leptoproduction observables.

\acknowledgments
It is a pleasure to thank the organizers of QCD-N12 for inviting me to well organized and 
interesting workshop.

\end{document}